\documentclass[twocolumn,showpacs,superscriptaddress,prl]{revtex4}
\usepackage[latin9]{inputenc}
\usepackage{color}
\usepackage{verbatim}
\usepackage{amsmath}
\usepackage{amssymb}
\usepackage{graphicx}
\usepackage{esint}
\usepackage{hyperref}

\makeatletter
\@ifundefined{textcolor}{}
{%
 \definecolor{BLACK}{gray}{0}
 \definecolor{WHITE}{gray}{1}
 \definecolor{RED}{rgb}{1,0,0}
 \definecolor{GREEN}{rgb}{0,1,0}
 \definecolor{BLUE}{rgb}{0,0,1}
 \definecolor{CYAN}{cmyk}{1,0,0,0}
 \definecolor{MAGENTA}{cmyk}{0,1,0,0}
 \definecolor{YELLOW}{cmyk}{0,0,1,0}
}

\usepackage{mathrsfs}

\draft
\makeatother

\begin{document}

\title{Detecting Topological Phases in Cold Atoms}

\author{Xiong-Jun Liu}
\affiliation{Department of Physics, Massachusetts Institute of Technology, Cambridge, Massachusetts 02139, USA}
\affiliation{Department of Physics, Hong Kong University of Science and Technology, Clear Water Bay, Hong Kong, China}
\affiliation{Institute for Advanced Study, Hong Kong University of Science and Technology, Clear Water Bay, Hong Kong, China}
\author{K. T. Law}
\affiliation{Department of Physics, Hong Kong University of Science and Technology, Clear Water Bay, Hong Kong, China}
\author{T. K. Ng}
\affiliation{Department of Physics, Hong Kong University of Science and Technology, Clear Water Bay, Hong Kong, China}
\author{Patrick A. Lee}
\affiliation{Department of Physics, Massachusetts Institute of Technology, Cambridge, Massachusetts 02139, USA}

\begin{abstract}
Chern insulators are band insulators which exhibit a gap in the bulk and gapless excitations in the edge. Detection of Chern insulators is a serious challenge in cold atoms since the Hall transport measurements are technically unrealistic for neutral atoms.
By establishing a natural correspondence between the time-reversal invariant topological insulator and quantum anomalous Hall system, we show for a class of Chern insulators that the topology can be determined by only measuring Bloch eigenstates at highly symmetric points of the
Brillouin zone (BZ).
Furthermore, we introduce two experimental schemes including the spin-resolved Bloch oscillation to
carry out
the measurement.
These schemes are highly feasible
under realistic experimental conditions. Our results may provide a powerful tool to detect topological phases in cold atoms.
\end{abstract}
\date{\today }
\maketitle

\indent

The recent great advancement in realizing synthetic Spin-orbit (SO) coupling \cite{Liu,Lin,Chapman,Pan,Wang,MIT} turns cold atom systems into new and promising platforms to probe exotic topological phases beyond natural conditions \cite{1DTI,Seo,Victor}.
So far the experimentally realized SO interaction \cite{Lin,Chapman,Pan,Wang,MIT} is a one-dimensional (1D) SO term with equal Rashba and Dresselhaus amplitudes through a two-photon Raman process as proposed by an earlier theoretical work~\cite{Liu}. The study of higher dimensional topological phases necessitates the realization of higher dimensional SO interactions in the cold atoms \cite{Liu2,Chuanwei1,Sato,Vyasanakere,BCS-BEC1,BCS-BEC2,Qi}. While a 2D or 3D synthetic SO term is yet to be realized in experiment, theoretical schemes have been
proposed \cite{Ruseckas,Chuanwei2,Ian}. In particular, it was proposed in a very recent work that the 2D SO interaction can be realized with realistic cold atom platforms \cite{LiuTSF}. In the single particle regime this model describes a quantum anomalous Hall (QAH) insulator (Chern insulator) \cite{Haldane} which exhibits a gap in the bulk and chiral edge modes in the boundary \cite{Wu,Liu1,Zhang,Hauke}.

Due to the absence of local orders, topological phases are typically hard to detect. For cold atoms the
task may be
even more demanding since quantized (Hall)
transport measurements~\cite{Chang},
widely exploited in the solid state systems, are technically unrealistic for neutral atoms.
On the other hand,
while the
detection of gapless edge modes by, e.g., light Bragg scattering proposed in~\cite{Liu1} is
in principle straightforward,
its usefulness may be limited
by the difficulty in separating edge state signals from the
bulk background and
complicated by realistic boundary conditions~\cite{Goldman,Buchhold,Goldman1}. Alternative strategies for the detection include to measure the state-dependent atom density response to external field~\cite{Shao}, the bulk Chern number from Berry curvature over the Brillouin zone (BZ)~\cite{Alba,Price}, Zak's phase~\cite{Demler}, and charge pumping~\cite{WangLei}. While these methods provide direct detection of the bulk topological invariants, they rely on complicated manipulations or measurements on the whole bulk band, which may still be challenging for the delicate cold atom systems.

In this letter, we propose to detect Chern insulators by measuring the bulk states at only few highly symmetric points of the BZ. By establishing a natural correspondence between the time-reversal (TR) invariant topological insulator and the QAH system, we show for a class of Chern insulators that the topology can be determined by measuring the Bloch eigenstates at only highly symmetric points of the BZ. This greatly simplifies measurement in the realistic experiments. We further introduce experimental schemes including the spin-resolved Bloch oscillation in 2D optical lattices to
carry out
the measurement of the topological states.

We consider the square optical lattice model for spin-$1/2$ cold atoms proposed in a recent work, which has essential advantages in
its experimental
realization~\cite{LiuTSF}. The main results of this letter, as shown below and detailed in the Supplementary Material~\cite{SI}, can be applied to more general Chern insulators and lattice configurations including honeycomb lattices. The Hamiltonian of the system $H=\sum_{\bold k,\sigma\sigma'}\hat c_{\bold k,\sigma}^{\dag}{\cal H}_{\sigma,\sigma'}(\bold k)\hat
c_{\bold k,\sigma'}$, with
\begin{eqnarray}\label{eqn:H1}
{\cal H}(\bold k)=d_x(\bold k)\sigma_x+d_y(\bold k)\sigma_y+d_z(\bold k)\sigma_z,
\end{eqnarray}
where $d_z=m_z-2t_0\cos(k_xa)-2t_0\cos(k_ya)$ and $d_{x,y}=-2t_{\rm so}\sin(k_{y,x}a)$, with
$m_z$ a controllable Zeeman splitting induced by a small two-photon off-resonance in the Raman couplings, $t_0$ and $t_{\rm so}^{(0)}$ representing the nearest-neighbor spin-conserved and spin-flipped hopping coefficients, respectively~\cite{LiuTSF}.
In the cold atom context, spin refers to two hyperfine levels.
The topology of the system is characterized by the first Chern number expressed by the integral of the Berry curvature over the first BZ: $C_1=\frac{1}{2\pi}\int dk_xdk_y{\cal B}_{-,z}(\bold k)$, where $\bold {\cal B}_\pm(\bold k)=\nabla_{\bold k}\times\bold{\cal A}_\pm(\bold k)$ and $\bold{\cal A}_\pm(\bold k)=i\hbar\langle u_{\pm,\bold k}|\nabla_{\bold k}|u_{\pm,\bold k}\rangle$, with $|u_{\pm,\bold k}\rangle$
labeling the upper and lower
 Bloch eigenstates solved from the Hamiltonian~\eqref{eqn:H1}. Direct calculation shows that $C_1=\mbox{sgn}(m_z)$ when $0<|m_z|<4t_0$, and otherwise $C_1=0$.

A Chern insulator explicitly breaks TR symmetry. Nevertheless, the above Hamiltonian $H$ is symmetric under the 2D inversion transformation defined by $P=\hat P\otimes\hat R_{\rm 2D}$, where $\hat P=\sigma_z$ acting on spin space and the 2D spatial operator $\hat R_{2D}$ transforms Bravais
lattice vector $\bold R\rightarrow-\bold R$. For the Bloch Hamiltonian we have $\hat P{\cal H}(\bold k)\hat P^{-1}={\cal H}(-\bold k)$, which follows that at the four highly symmetric points $[\hat P,{\cal H}(\bold \Lambda_i)]=0$, with $\{\bold \Lambda_i\}=\{(0,0), (0,\pi), (\pi,0), (\pi,\pi)\}$. Therefore the Bloch states $|u_{\pm}(\bold \Lambda_i)\rangle$ are also eigenstates of the parity operator $\hat P$, with eigenvalues $\xi^{(\pm)}_i=+1$ or $-1$. Similar as in topological insulators \cite{Fu1,Fu2}, we define the following invariant
\begin{eqnarray}\label{eqn:parity}
(-1)^\nu=\prod_i\xi^{(-)}(\bold \Lambda_i),
\end{eqnarray}
and can verify
for
Hamiltonian~\eqref{eqn:H1} that $\nu=0$ when the system is in the trivial regime, and $\nu=1$ for the topological regime.
Furthermore,
for the present square lattice which has four Dirac points coinciding with the four highly symmetric points, the Chern number
is given by
\begin{eqnarray}\label{eqn:Chern}
C_1=-\frac{\nu}{2}\sum_i\xi^{(-)}(\bold \Lambda_i).
\end{eqnarray}
It is straightforward to check that when the Zeeman term varies from $m_z\gtrsim0$ to $m_z\lesssim0$, two parity eigenvalues $\xi^{(-)}_{\bold \Lambda_2,\bold \Lambda_3}$ change sign and then $C_1$ changes from $1$ to $-1$.

For a more general Hamiltonian with inversion symmetry, we demonstrate below and show in more detail in the Supplementary Material \cite{SI} that when $\nu$ given by Eq.~\eqref{eqn:parity} is 1, the Chern number is odd and the system is nontrivial, while for $\nu=0$, the number $C_1$ is even and the system may or may not be nontrivial. The essential idea is to adopt the topological classification of a TR variant topological insulator which can be artificially constructed from the studied Chern insulator and its time-reversed artificial copy. Following Fu-Kane's theorem (see Supplementary Material for more details \cite{SI}), the topological invariant of a 2D TR invariant topological insulator with inversion symmetry can be determined by the product of the four parity eigenvalues as given in Eq.~\eqref{eqn:parity} \cite{Fu1,Fu2}. Note that the TR invariant topological insulator is constructed by two independent time-reversed copies of Chern insulators. Generically, the {\it trivial} $(\nu=0)$ and {\it topological} ($\nu=1$) phases of the TR invariant topological insulator respectively correspond to the {\it even} and {\it odd} Chern numbers for the two copies of QAH insulators. Therefore, the invariant given by Eq.~\eqref{eqn:parity} exactly describes the topology of a Chern insulator when $|C_1|$ is either $0$ or $1$, which is true for most of the available theoretical models in cold atoms \cite{Wu,Liu1,Zhang,Hauke,Goldman,Buchhold,Goldman1,Shao,Alba,Price,Demler}.

The formula~\eqref{eqn:parity} is the central result of this work to be applied to the detection of Chern insulators which preserve parity symmetry and satisfy $|C_1|=\{0,1\}$. Note that for a cold atom system the possible values of $C_1$ can be exactly known by theory. Since only the four parity eigenstates at $\bold k=\bold\Lambda_i$ need to be measured, the procedure of detecting a Chern insulator can be essentially simplified. In the rest of this letter we study two different approaches based on Bloch oscillation \cite{Bloch,Tarruell} to detect the Chern insulating phases given by the Hamiltonian~\eqref{eqn:H1}. Note that Eq.~\eqref{eqn:parity} can apply to other lattice configurations such as honeycomb lattices, with new results predicted~\cite{SI}. Also, these results are valid for both fermions and bosons trapped in the 2D optical lattice.

First we consider to measure the topological invariants from the Berry curvature $\bold {\cal B}_+=-\bold {\cal B}_-$ at the four symmetric points. A straightforward calculation yields that
\begin{eqnarray}\label{eqn:Berry1}
\bold {\cal B}_-=\bigr[\frac{2m_zt_{\rm so}^2}{d^3(\bold k)}\cos k_x\cos k_y-\frac{4t_st_{\rm so}^2}{d^3(\bold k)}(\cos k_x+\cos k_y)\bigr]\hat e_z
\end{eqnarray}
with $d(\bold k)=[\sum_jd_j^2(\bold k)]^{1/2}$. It can be verified that $\mbox{sgn}[{\cal B}_{-,z}(0,\pi)]=\mbox{sgn}[{\cal B}_{-,z}(\pi,0)]=\mbox{sgn}[\xi^{(-)}_{0,\pi}]=\mbox{sgn}[\xi^{(-)}_{\pi,0}]$, and $\mbox{sgn}[{\cal B}_{-,z}(0,0)]=-\mbox{sgn}[{\cal B}_{-,z}(\pi,\pi)]=-\mbox{sgn}[\xi^{(-)}_{0,0}]=\mbox{sgn}[\xi^{(-)}_{\pi,\pi}]$, which follows that $\prod_i\mbox{sgn}[{\cal B}_{-,z}(\bold\Lambda_i)]=\prod_i\xi^{(-)}(\bold \Lambda_i)$. Therefore,
measuring $\xi^{(-)}(\bold \Lambda_i)$ reduces to
measuring $\mbox{sgn}({\cal B}_{-,z})$ at the four momenta $\bold k=\bold\Lambda_i$, which can be carried out by Bloch oscillation. We emphasize that the present approach is essentially different from that in Ref.~\cite{Price} which suggests to detect the bulk Chern number by measuring quantitatively $\bold{\cal B}_{-}$ over the BZ. Here only the signs of ${\cal B}_{-,z}$ at four points need to be measured. Note that we always have ${\cal B}^-_{0,\pi}={\cal B}^-_{\pi,0}$. The measurement can be further simplified to determine $\bold {\cal B}_-$ at $\bold\Lambda_1=(0,0)$ and $\bold\Lambda_4=(\pi,\pi)$. In the presence of an external force $\bold F=F_x\vec e_x+F_y\vec e_y$ in the $x$-$y$ plane, which can be set by ramping the optical trapping or chirping the lattice frequency \cite{Bloch}, the Bloch wave packet oscillates along the direction of the force, while deflects to the transverse direction due to the Berry curvature. The deflection direction provides a straightforward measurement of the sign of the Berry curvature.

The semiclassical dynamics of a wave packet at band $n$ ($=\pm$) with center-of-mass position $\bold r_c$ and momentum $\bold k_c$ is given by $\hbar\dot{\bold k}_c=\bold F(\bold r_c), \dot{\bold r}_c=\bold v_{0n}-\bold F\times\bold {\cal B}_n(\bold k_c)$,
where $\bold v_{0n}=\hbar^{-1}\partial_{\bold k}{\cal E}_n(\bold k)$ and ${\cal E}_n(\bold k)$ is the dispersion relation. The second term in the equation of $\dot{\bold r}_c$ represents the anomalous velocity, leading to the transverse shift. Note that the transverse shift in a unit Bloch time
\begin{eqnarray}\label{eqn:shift1}
\Delta\bold r_c=-\int_0^{2\pi}d\bold k_c\times\bold {\cal B}_n(\bold k_c) \nonumber
\end{eqnarray}
is independent of the force strength, but sensitive to the magnitude of Berry curvature. To make the results easily distinguishable in realistic experiments an appreciable transverse shift is preferred, which can be achieved by tuning $m_z$ close to the phase transition points.
In particular, to measure ${\cal B}_{0,0}$ and ${\cal B}_{\pi,\pi}$, we can tune $m_z$ close to $4t_0$ and $-4t_0$, in which cases the transverse shift is dominated by the Berry curvature around $\bold\Lambda_1$ and $\bold\Lambda_4$, respectively. The signs of ${\cal B}_{0,0}$ and ${\cal B}_{\pi,\pi}$ are directly read out from directions of the transverse shift.

In the realistic experiment, one shall consider a cloud of cold atoms which are initially trapped by a harmonic optical potential and centered at $\bold r_0=0$. The square lattice potentials are adiabatically switched on, along with the switch-off of the harmonic trap, and the atoms populate the states at the band bottom (for bosons) or starting from the band bottom (for fermions) \cite{Bloch}. Applying the static force $\bold F$ in the 2D plane accelerates the atomic cloud, with the dynamics described by the distribution function $\rho_\pm(\bold r,\bold k,T,\tau)$, where $T$ is the temperature. The initial profile of the atomic cloud, determined by $\rho_\pm(\bold r,\bold k,T,0)$, can be adjusted by the optical trapping potentials before switching on the square lattice. For the case of a weak force satisfying that $\pi E_g^2\gg2at_{\rm so}|\bold F|$ with $E_g$ the band gap, we can neglect the Landau-Zener (LZ) transition between the lower to upper subbands \cite{LZ}. The evolution of $\rho_\pm(\bold r,\bold k,T,\tau)$, governed by the semiclassical Boltzmann equation, satisfies the ballistic law and reads
\begin{eqnarray}\label{eqn:Bloch2}
\rho_\pm(\bold r,\bold k,T,\tau)&=&\rho_\pm(\bold r-\int_0^\tau\bold v_{0\pm}dt'+\int_0^\tau\bold F\times\bold {\cal B}_\pm(\bold k)dt',\nonumber\\
&&\bold k-\int_0^\tau dt'\bold F/\hbar,T,0).
\end{eqnarray}
With this solution the dynamics of the atomic cloud can be studied numerically, as presented below.

\begin{figure}[t]
\includegraphics[width=1\columnwidth]{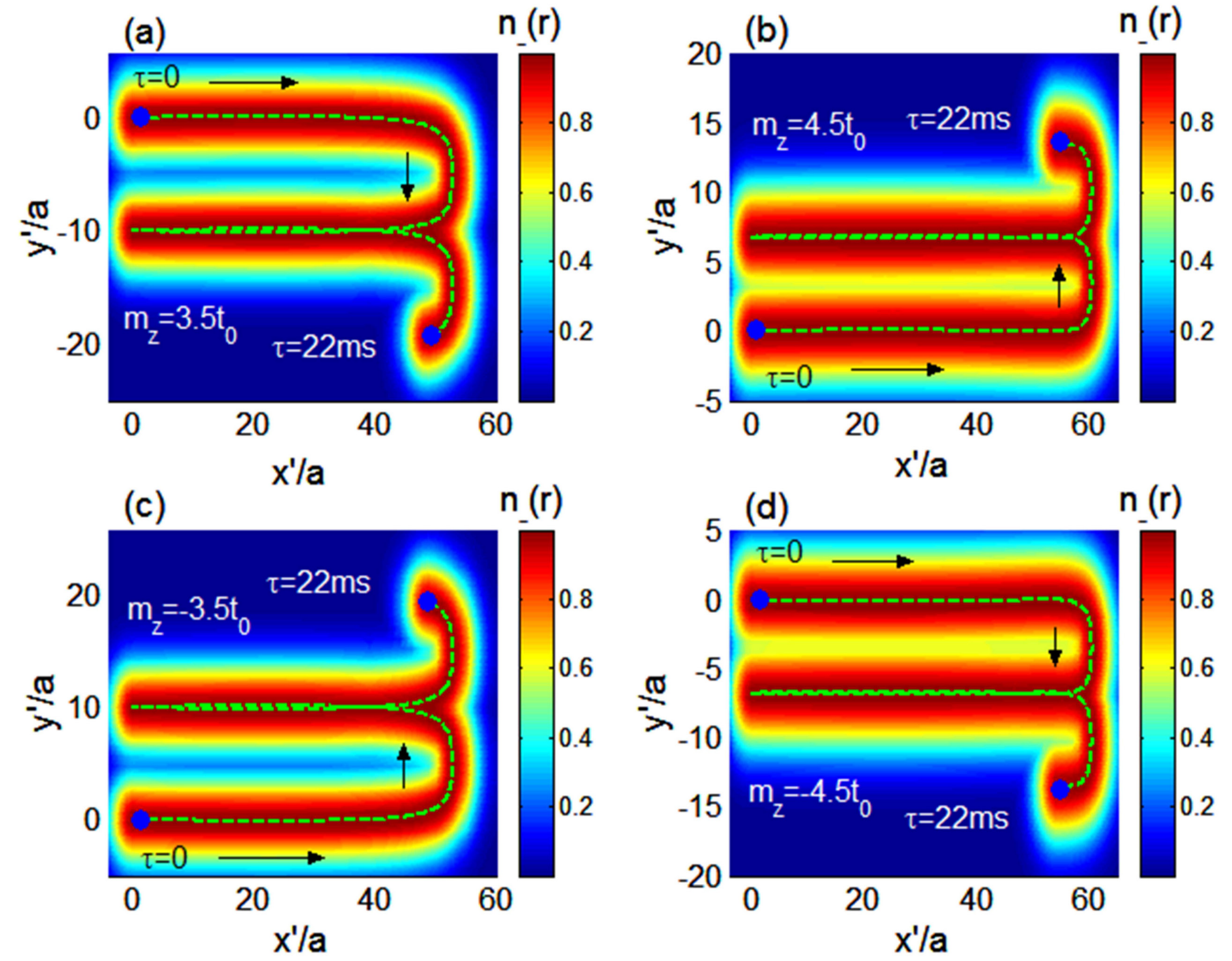}\caption{(Color online) Measuring $\mbox{sgn}({\cal B}_{-,z})$ at $\bold \Lambda_1=(0,0)$ in (a,b) and at $\bold \Lambda_4=(\pi,\pi)$ in (c,d) by Bloch oscillation, which determine the topology of the bulk band. The static force, applied along $\hat e_{x'}=(\hat e_x+\hat e_y)/\sqrt{2}$ direction, induces transverse deflection of the atomic cloud to $\pm\hat e_{y'}=\pm(\hat e_y-\hat e_x)/\sqrt{2}$. Other parameters are taken as $t_{\rm so}=t_0=2\pi\times0.52$kHz \cite{LiuTSF} and $|\bold F|=0.13t_0/a$.}
\label{Bloch1}
\end{figure}
The Berry curvatures ${\cal B}_{0,0}$ and ${\cal B}_{\pi,\pi}$ are determined from the numerical simulation in Fig.~\ref{Bloch1}, where we plot the atomic density at zero temperature $n_-(\bold r,\tau)=\int d^2\bold k\rho_-(\bold r,\bold k,0,\tau)$ with its maximum magnitude rescaled to be unit. Note that the band bottom of the lower subband is located at $\bold k_{\rm bot}=(\pi,\pi)$ for $m_z>0$, and at $\bold k_{\rm bot}=(0,0)$ if $m_z<0$. By applying $\bold F$ in the $\hat e_{x'}=(\hat e_x+\hat e_y)/\sqrt{2}$ direction, the atomic cloud oscillates in the position space along this direction while deflects to the transverse $\pm\hat e_{y'}=\pm(\hat e_y-\hat e_x)/\sqrt{2}$ direction due to the Berry curvature. By tuning the Zeeman parameter $m_z$ from less than to greater than $4t_0$, we can see that the transverse motion changes from $-\hat e_{y'}$ to $+\hat e_{y'}$ direction [Fig.~\ref{Bloch1}(a,b)], which implies ${\cal B}_{0,0}$ changes direction from $-z$ to $+z$ direction. On the other hand,
by tuning $m_z$ from $m_z>-4t_0$ to $m_z<-4t_0$, the transverse motion changes from $+\hat e_{y'}$ to $-\hat e_{y'}$ direction, again implying that ${\cal B}_{\pi,\pi}$ changes from $+z$ to $-z$ direction [Fig.~\ref{Bloch1}(c,d)]. With these results one gets that $\nu=1$ for $0<|m_z|<4t_0$ and $\nu=0$ for $|m_z|\geq4t_0$, and the topological phase is obtained in the former regime. Similar phenomena can be obtained for Berry curvatures ${\cal B}_{\bold\Lambda_2}$ and ${\cal B}_{\bold\Lambda_3}$ by tuning $m_z$ close to zero (not shown here).
In this way we further find that $C_1=\mbox{sgn}(m_z)$ in the topological phase.

Now we introduce another approach for the measurement: the spin-resolved Bloch oscillation. We shall determine the topology of the bulk by measuring the spin-polarization of the atomic cloud at the highly symmetric points. Note for the present square lattice the parity operator $\hat P=\sigma_z$, so the parity eigenstates are simply the spin eigenstates, with the spin-up and spin-down corresponding to different hyperfine levels. It then follows that $\prod_i\xi^{(-)}(\bold \Lambda_i)=\prod_i\mbox{sgn}[p_{\rm spin}(\bold\Lambda_i)]$, where the spin-polarization density $p_{\rm spin}=(n_\uparrow-n_\downarrow)/n_-$ with $n_{\uparrow/\downarrow}$ the spin-up/-down component of the atomic density and $n_-=n_\uparrow+n_\downarrow$. The spin-polarization can be measured directly in experiment by imagining the densities of atoms in two different hyperfine levels. Similarly, since the polarizations for the states at $\bold k=\bold\Lambda_2$ and $\bold k=\bold\Lambda_3$ are the same, the bulk topology is determined by $\mbox{sgn}(p_{\rm spin})$ at $\bold k=\bold\Lambda_1$ and $\bold k=\bold\Lambda_4$.

\begin{figure}[ht]
\includegraphics[width=1.0\columnwidth]{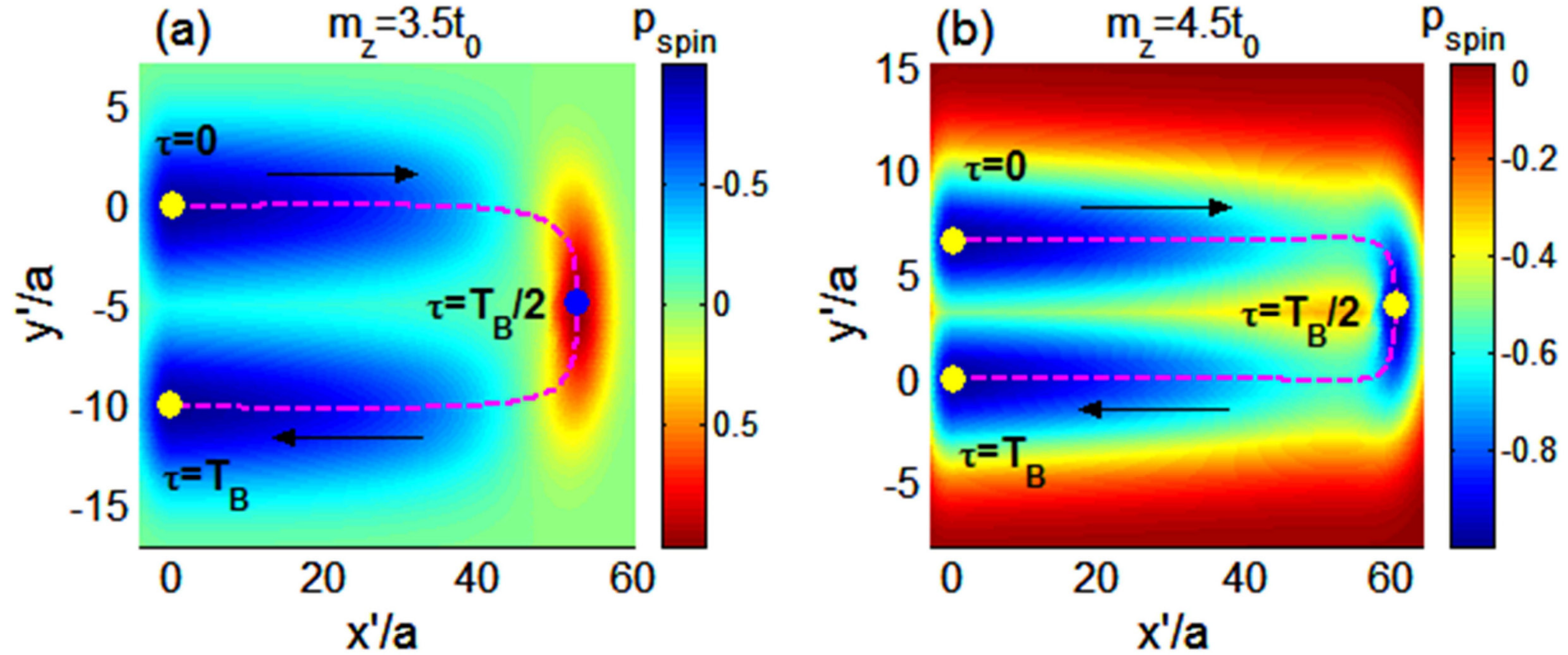}\caption{(Color online) Spin-resolved Bloch Oscillation with the force along $\hat e_{x'}$ direction. (a) Topological regime with $m_z=3.5t_0$. The spin-polarization $\mbox{sgn}[p_{\rm spin}(0)]=-\mbox{sgn}[p_{\rm spin}(T_B/2)]$; (b) Trivial regime with $m_z=4.5t_0$. The spin-polarization $\mbox{sgn}[p_{\rm spin}(0)]=\mbox{sgn}[p_{\rm spin}(T_B/2)]$.}
\label{spinBloch}
\end{figure}
In Fig.~\ref{spinBloch} we numerically plot the spin-polarization density $p_{\rm spin}(\tau)$ in the Bloch oscillation with $\bold F$ along $\hat e_{x'}$ direction. For the case $0<m_z<4t_0$, the spin-polarization of the atomic cloud, starting with the center momentum $\bold k=(0,0)$, changes from $p_{\rm spin}\simeq-1$ at $\tau=0$ to $p_{\rm spin}\simeq1$ at half Bloch time $\tau=T_B/2$, and then reverses back to $p_{\rm spin}\simeq-1$ again when a unit Bloch period is finished at $\tau=T_B$ [Fig.~\ref{spinBloch}(a)]. The sign change of the spin-polarization in a unit Bloch oscillation tells
us that $\nu=1$ and the mass terms exhibit opposite signs for the Dirac equations around $\bold\Lambda_1=(0,0)$ and $\bold\Lambda_4=(\pi,\pi)$, so the system is in topological phase. Fig.~\ref{spinBloch}(b) shows that in the case $m_z\geq4t_s$ the spin-polarization
$p_{\rm spin}<0$ for the whole Bloch oscillation, and therefore the phase is trivial.
The results for $m_z<0$ are similar. With these phenomena one can again determine that the topological phase is obtained when $0<|m_z|<4t_0$, with the Chern number $C_1=\mbox{sgn}(m_z)$. It is noteworthy that the spin-resolved Bloch oscillation does not 
require a large transverse shift in the Bloch oscillation, and thus can be more straightforward for experimental studies.

\begin{figure}[b]
\includegraphics[width=1.0\columnwidth]{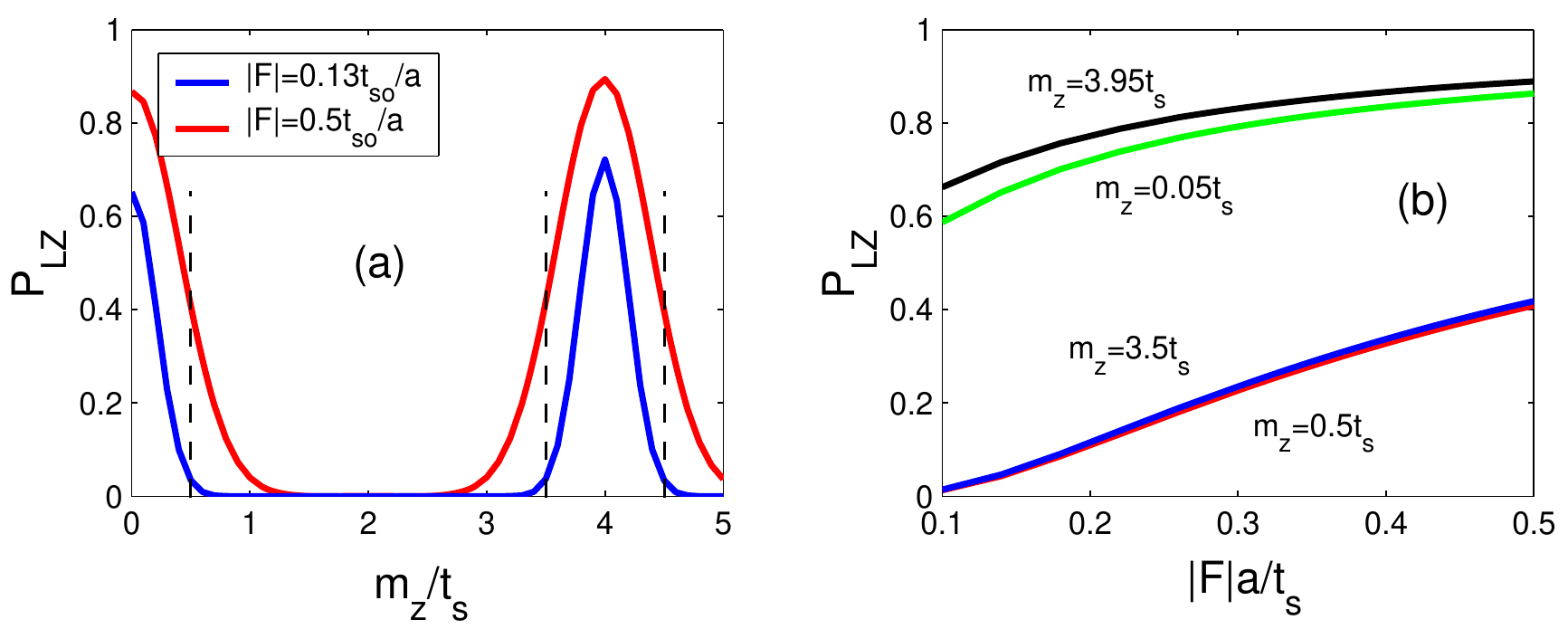}\caption{(Color online) Landau Zener probability $P_{\rm LZ}$ for the atomic cloud evolving through the avoided band crossing point, as a function of $m_z$ (a), and of $|\bold Fa|$ (b). For the numerical calculation the initial atomic cloud has a narrow momentum distribution with radius relative to the central momentum (at band bottom) about $k_{\rm rad}\simeq0.2/a$.}
\label{Zenerprobility}
\end{figure}
So far we have considered the weak static force regime. In the opposite parameter regime with $\pi E_g^2<2at_{\rm so}|\bold F|$, the LZ transition between the lower and higher subbands must be taken into account when the center-of-mass momentum of the atomic cloud is accelerated to the avoided crossing point \cite{LZ}. For convenience, we denote that $\bold k=(k_\perp,k_\parallel)$ with $k_\perp$ and $k_\parallel$ the momenta perpendicular and parallel to $\bold F$, respectively. Since $k_\perp$ is unchanged, it is useful to define $\tilde{E}_g(k_\perp)=\min[{\cal E}_+(k_\perp)]$ as the minimum of the upper subband energy with fixed $k_\perp$. Then a state initially at lower band and with momentum $\bold k$ transitions to the upper band with
the LZ probability $\bar p(k_\perp)=e^{-\pi\tilde{E}^2_g(k_\perp)/(v_F|\bold F|)}$~\cite{LZ}.
The averaging transition probability is given by $P_{\rm LZ}=\int d^2\bold r d^2\bold k\rho_-(\bold r,\bold k)\bar p(k_\perp)/\int d^2\bold r n_-(\bold r)$, with numerical results shown in Fig.~\ref{Zenerprobility} for different parameter regimes.

\begin{figure}[t]
\includegraphics[width=1.0\columnwidth]{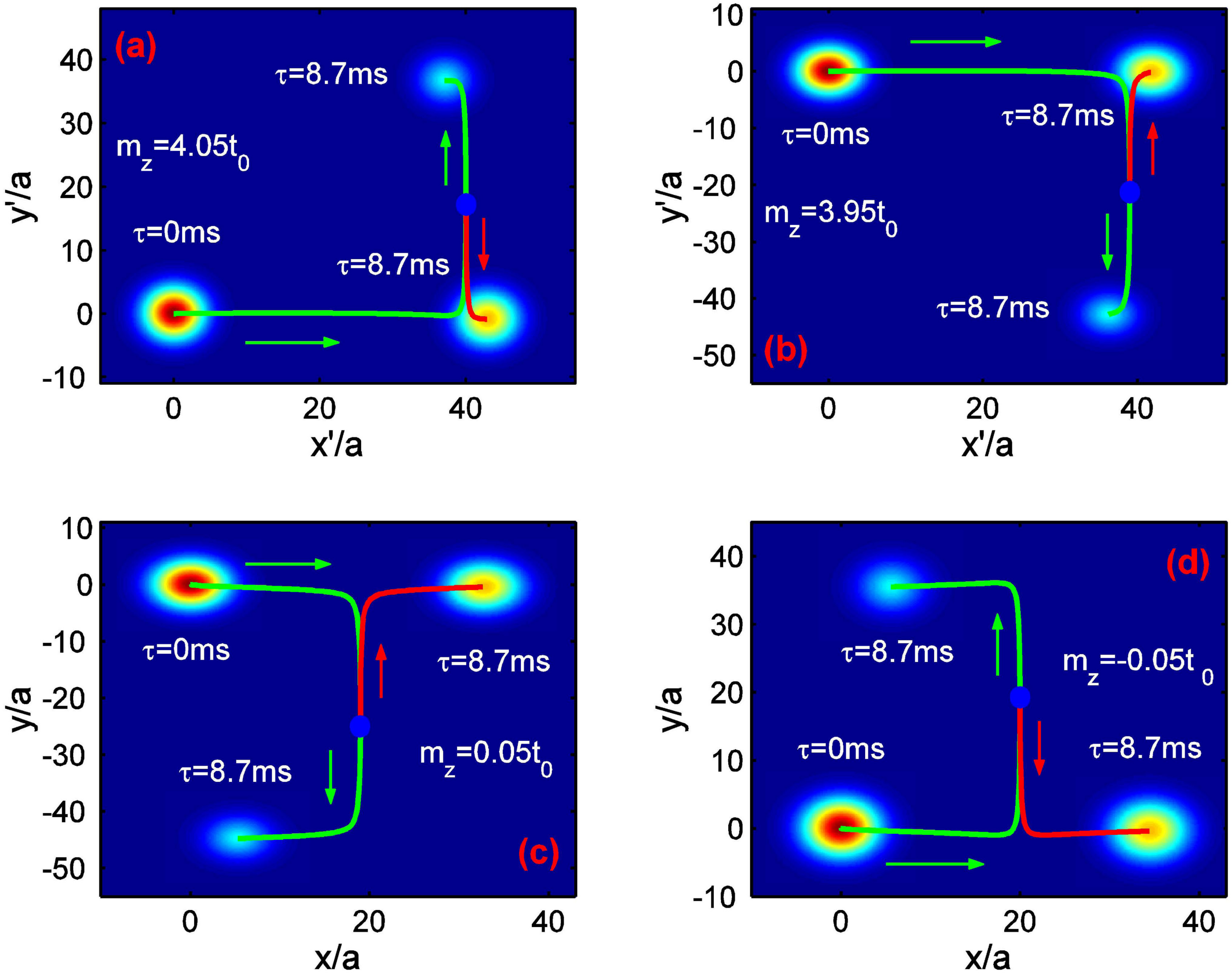}\caption{(Color online) Atomic cloud splitting induced by LZ transition close to phase transition points. A tiny bulk gap opens at $\bold k=\bold\Lambda_1$ for $m_z=4.05t_0$ (a) and $m_z=3.95t_0$ (b), and at $\bold k=\bold\Lambda_2,\bold\Lambda_3$ for $m_z=0.05t_0$ (c) and $m_z=-0.05t_0$ (d), with the magnitude of the gap $E_g=0.05t_0$. Other parameters are taken as $t_{\rm so}=t_0=2\pi\times0.52$kHz, and the force $|\bold F|=0.2t_0/a$ applied along $\hat e_{x'}$ (a,b) and $\hat e_x$ (c,d) directions, respectively. The arrows represent the directions of atomic cloud motion.}
\label{Zener}
\end{figure}
The LZ transition leads to the splitting of the atomic cloud from one to two, which provides a detection of the critical point of the topological phase transition. The numerical results are shown in Fig.~\ref{Zener}. For the cases with $m_z\gtrsim4t_s$ [Fig.~\ref{Zener}(a)] and $m_z\lesssim4t_s$ [Fig.~\ref{Zener}(b)], a tiny bulk gap is located at $\bold\Lambda_1=(0,0)$. Applying the static force along $\hat e_{x'}$ direction we can see that the atomic cloud first is deflected to $\hat e_{y'}$ [for (a)] or $-\hat e_{y'}$ [for (b)] direction, and then splits into two when the central momentum is accelerated to the $\bold\Lambda_1$ point. The atomic cloud corresponding to the lower subband keeps the original deflection route, while the new cloud formed by the atoms in the upper band states is deflected oppositely, reflecting that ${\cal B}_\pm$ are opposite in the $z$ direction. Similarly, by tuning $m_z\sim0$, one can detect the phase transition due to the gap closing at $\bold\Lambda_2$ and $\bold\Lambda_3$ by applying the force along $\hat e_x$ (or $\hat e_y$) direction [Fig.~\ref{Zener}(c,d)].

In conclusion we
proposed to detect 2D Chern insulators by measuring the bulk states at four highly symmetric points of the BZ. From a natural correspondence between the TR invariant topological insulator and the QAH system, we show for a class of Chern insulators that the topology can be determined by measuring the parity eigenstates at only highly symmetric points of the BZ. Moreover, the detection relies on only qualitative rather than quantitative measurements on physical numbers. This enables a much simpler strategy to detect the topological phases comparing with conventional methods to measure the edge states or bulk Chern invariants. We further introduced two realistic experimental schemes to carry out the measurement, and our detection strategy can be applied to both square and honeycomb lattice systems, the two most relevant configurations for the cold atom experiments~\cite{SI}. It is noteworthy that these schemes can also be directly applied to the detection of TR invariant topological insulators. Our work showcases the advantages of cold atoms since the parity eigenstates are hard to directly measure in condensed matter system. Our results can provide a powerful tool to detect topological phases in cold atoms.

We thank Waseem Bakr, Lawrence W. Cheuk, Wujie Huang, Tin-Lun Ho, Liang Fu, and Junwei Liu
for very helpful discussions. We acknowledge the support from HKRGC through Grant 605512 and HKUST3/CRF09. PAL acknowledges the support by DOE Grant DE-FG-02-03-ER46076.


\noindent

\noindent

\onecolumngrid

\renewcommand{\thesection}{S-\arabic{section}}
\renewcommand{\theequation}{S\arabic{equation}}
\setcounter{equation}{0}  
\renewcommand{\thefigure}{S\arabic{figure}}
\setcounter{figure}{0}  

\section*{\Large\bf Supplementary Information}

In this Supplementary Material we provide the details for the generic proof of topological invariant for Chern insulators with inversion symmetry, and the discussion on applying our detection strategy to honeycomb lattice system. Interesting new results are found in the honeycomb lattice system.

\section{Topological Invariant for Chern insulators}

Here we provide the details of proving Eq.~(2) in the main text for more generic Chern insulators, following the theory by Fu and Kane~\cite{SIFu1,SIFu2}. For the proof we consider a multi-band Chern insulator described the Hamiltonian $H=\sum_{\bold k,\sigma\sigma'}\hat c_{\bold k,\sigma}^{\dag}{\cal H}_{\sigma,\sigma'}(\bold k)\hat
c_{\bold k,\sigma'}$, which is invariant under a 2D inversion symmetry defined by $P=\hat P\otimes\hat R_{2D}$. Here the 2D spatial operator $\hat R_{2D}$ transforms Bravais lattice vector $\bold R\rightarrow-\bold R$, and $\hat P$ acts on the spin (or pseudospin) space. For the two-band model considered in the main text $\hat P=\sigma_z$. The inversion symmetry $P{\cal H}(\bold k) P^{-1}={\cal H}(\bold k)$ implies that
\begin{eqnarray}\label{eqn:invariant1}
\hat P{\cal H}(\bold k)\hat P^{-1}={\cal H}(-\bold k).
\end{eqnarray}
Then at the four symmetric points $\{\bold \Lambda_i\}=\{(0,0), (0,\pi), (\pi,0), (\pi,\pi)\}$ the Bloch Hamiltonian ${\cal H}(\bold k)$ is invariant under parity transformation $\hat P$.

To construct a time-reversal (TR) invariant topological insulator we introduce an artificial TR operator $\Theta=i\tau_yK$, with $K$ the complex conjugation and $\tau_y$ a Pauli matrix acting on the artificial spin space. The considered Chern insulator is assumed to be in the spin-up state of the operator $\tau_z$. Using the TR operator the time-reversed copy of ${\cal H}(\bold k)$ can be created by $\tilde{\cal H}(-\bold k)=\Theta{\cal H}(\bold k)\Theta^{-1}$. With the two copies a TR invariant topological insulator can be constructed with the Hamiltonian
\begin{eqnarray}\label{eqn:invariant2}
{\cal H}_{\rm TR}={\cal H}(\bold k)\oplus\tilde{\cal H}(-\bold k).
\end{eqnarray}
According to Fu-Kane's theorem, the topological invariant of a TR invariant topological insulator can be determined by Pfaffian of the antisymmetric matrix of TR operator $w_{mn}=\langle u_m(-\bold k)|\Theta|u_n(\bold k)\rangle$ at the four TR invariant momenta $\bold k=\bold\Lambda_i$, where $n,m$ are subband indices. The antisymmetry of the matrix $w_{mn}=-w_{nm}$ can be verified using the property $\Theta^2=-1$. The invariant is given by~\cite{SIFu1}
\begin{eqnarray}\label{eqn:invariant3}
(-1)^{\nu}=\prod_i\frac{\sqrt{\det[w(\bold\Lambda_i)]}}{{\rm Pf}[w(\bold\Lambda_i)]}.
\end{eqnarray}
When the system preserves inversion symmetry, one can define another antisymmetric matrix by
\begin{eqnarray}\label{eqn:invariant4}
v_{mn}=\langle u_m(\bold k)|P\Theta|u_n(\bold k)\rangle.
\end{eqnarray}
Note that $P^2=1$ we have $w_{mn}=\langle u_m(-\bold k)|P(P\Theta)|u_n(\bold k)\rangle$. Since $\hat R_{2D}\bold\Lambda_i\hat R_{2D}^{-1}=\bold\Lambda_i$ and $|u_n(\bold \Lambda_i)\rangle$ are also parity eigenstates of the operator $\hat P$, we have $w_{mn}(\bold\Lambda_i)=\xi_m(\bold\Lambda_i)v_{mn}(\bold\Lambda_i)$. The Pfaffian of the matrix $w_{mn}$ at the symmetric points takes the form~\cite{SIFu2}
\begin{eqnarray}\label{eqn:invariant5}
{\rm Pf}[w(\bold\Lambda_i)]={\rm Pf}[v(\bold\Lambda_i)]\prod_{m=1}^N\xi_{2m}^{(-)}(\bold \Lambda_i),
\end{eqnarray}
where $\xi_{2m}^{(-)}(\bold \Lambda_i)$ are parity eigenvalues of the states at the $N$ occupied subbands corresponding to the Hamiltonian ${\cal H}(\bold k)$. We have assumed that $|u_{2n+1}(\bold k)\rangle$ are the eigenstates of the Hamiltonian $\tilde{\cal H}(-\bold k)$. For a system with both TR symmetry and inversion symmetry, it can be shown that ${\rm Pf}[v(\bold\Lambda_i)]=1$~\cite{SIFu2}, and it then follows that
\begin{eqnarray}\label{eqn:invariant6}
(-1)^{\nu}=\prod_{i=1}^4\prod_{m=1}^N\xi_{2m}^{(-)}(\bold \Lambda_i).
\end{eqnarray}
For the two-band model, with only the lower subband occupied, the above formula reduces to the Eq.~(2) in the main text by taking $N=1$. It is important that while the TR invariant topological insulator is composed of two copies of Chern insulators, its topological invariant can be expressed by the product of parity eigenvalues in one single Chern insulator. This is the reason why for the inversion symmetric Chern insulators the topological phases may also be characterized by the parity eigenvalues.

Note that the TR invariant topological insulator is constructed by two independent time-reversed copies of Chern insulators. The topological phase of ${\cal H}_{\rm TR}$ requires that each copy of the Chern insulator be in the topological phase. Furthermore, since the {\it trivial} $(\nu=0)$ and {\it topological} ($\nu=1$) phases of a TR invariant topological insulator respectively correspond to the {\it even} and {\it odd} Chern numbers for the two copies of QAH insulators, the invariant given by Eq.~\eqref{eqn:invariant6} exactly describes the topology of a Chern insulator only when $|C_1|$ is either $0$ or $1$, which is true for most of the available theoretical models in cold atoms.

\section{Honeycomb lattice model}

\begin{figure}[b]
\includegraphics[width=0.4\columnwidth]{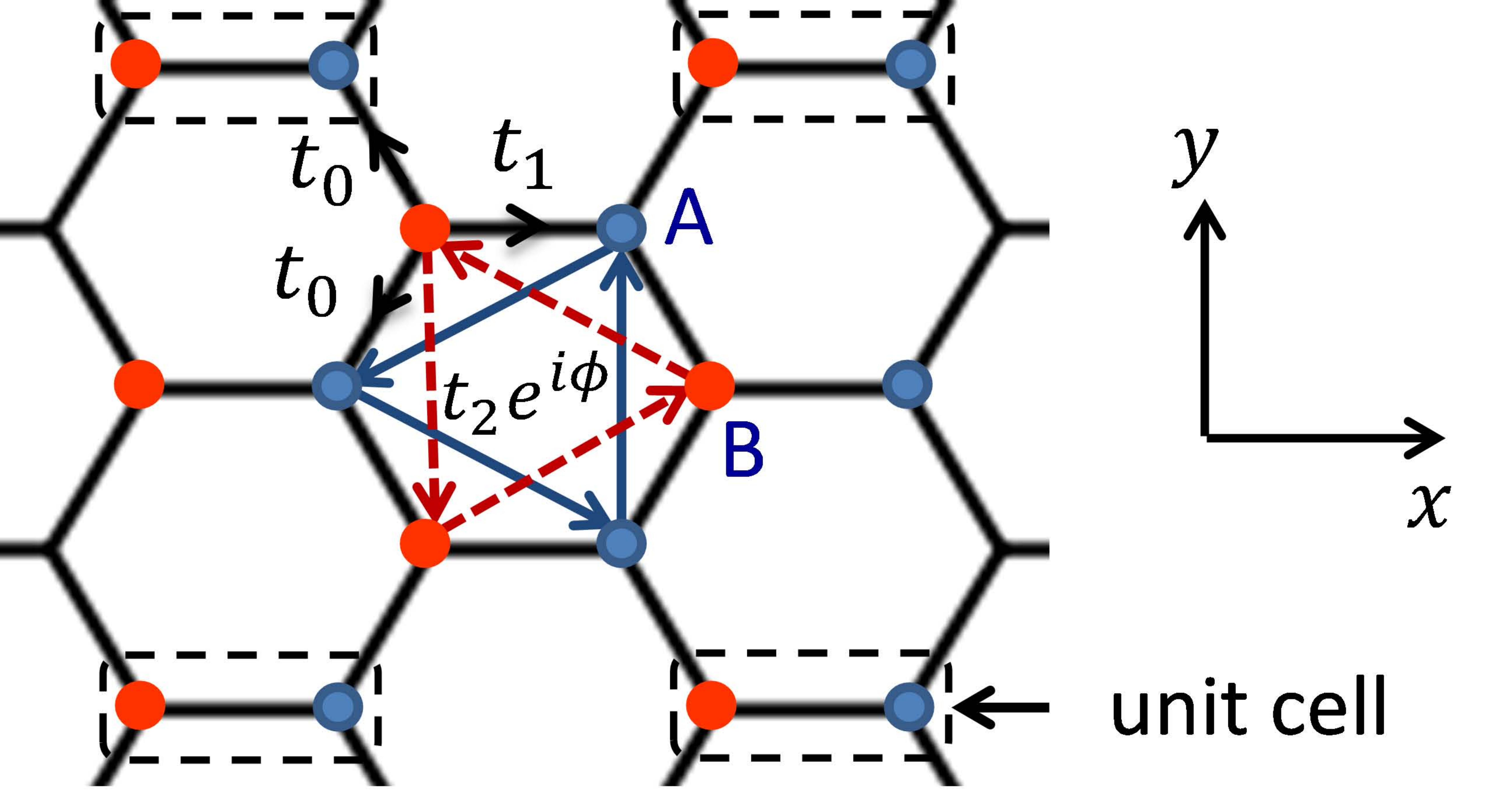} \caption{(Color online) Honeycomb lattice model for Chern insulating phase, with anisotropic nearest-neighbor hopping terms.}
\label{lattice}
\end{figure}
As first proposed by Haldane, in the presence of complex next-nearest-neighbor hopping and with the hopping phases induced by staggered flux, the honeycomb lattice can be driven into quantum anomalous Hall insulating phase~\cite{SIHaldane}. Here we consider a generalized Haldane model with anisotropic hopping terms illustrated in Fig.~\ref{lattice}. The nearest-neighbor hopping along the horizontal ($x$) direction is different from those along the other two directions $t_1\neq t_0$. This configuration can be easily achieved by varying the barrier height for the horizontal bond or the bond length in the honeycomb lattice, which can be controlled by tuning intensities of lasers used to generate the lattice, and was realized in a recent experiment~\cite{SITarruell}. In cold atoms the hopping phases for the next-nearest-neighbor hopping terms~\cite{SIHaldane} may be generated by inducing a staggered gauge potential~\cite{SIShao}. The anisotropy in the magnitudes of the next-nearest-neighbor hopping terms does not affect the phase diagram shown below, and therefore we neglect it for the present study. Properly choosing the unit cell as shown in Fig.~\ref{lattice}~\cite{SIMermin} and from the tight-binding model, we obtain the Hamiltonian $H=\sum_{\bold k,\sigma\sigma'}\hat c_{\bold k,\sigma}^{\dag}{\cal H}_{\sigma,\sigma'}(\bold k)\hat
c_{\bold k,\sigma'}$, with
\begin{eqnarray}\label{eqn:SIH1}
{\cal H}(\bold k)&=&\bigr[t_1+t_0\cos(\frac{3}{2}k_xa+\frac{\sqrt{3}}{2}k_ya)+t_0\cos(\frac{3}{2}k_xa-\frac{\sqrt{3}}{2}k_ya)\bigr]\sigma_x\nonumber\\
&&+t_0\bigr[\sin(\frac{3}{2}k_xa+\frac{\sqrt{3}}{2}k_ya)+\sin(\frac{3}{2}k_xa-\frac{\sqrt{3}}{2}k_ya)\bigr]\sigma_y\nonumber\\
&&+\biggr\{m_z+2t_2\sin\phi\bigr[\sin(\frac{3}{2}k_xa+\frac{\sqrt{3}}{2}k_ya)-\sin(\frac{3}{2}k_xa-\frac{\sqrt{3}}{2}k_ya)-\sin(\sqrt{3}k_ya)\bigr]\biggr\}\sigma_z.
\end{eqnarray}
Here the Pauli matrices $\sigma_{x,y,z}$ act on the sublattice space, $\phi$ is the next-nearest-neighbor hopping phase, and $m_z$ represents an onsite energy difference between A and B sublattice sites. In the isotropic case with $t_1=t_0$, the above Hamiltonian reduces to standard Haldane model with the topological phase obtained in the parameter regime~\cite{SIHaldane}: $|m_z|<3\sqrt{3}t_2|\sin\phi|$, and the Chern number reads $C_1=\mbox{sgn}(\sin\phi)$ (assuming $t_0,t_2>0$). For the anisotropic case, in the following we predict a new topological phase transition governed by the difference between $t_0$ and $t_1$. In this work we are interested in the latter case with zero Zeeman term and study the topological phase transition by tuning $t_1$, which is feasible for the realistic experiments~\cite{SITarruell}.

For convenience, we denote that $k_1=\frac{3}{2}k_xa+\frac{\sqrt{3}}{2}k_ya$ and $k_2=\frac{3}{2}k_xa-\frac{\sqrt{3}}{2}k_ya$. The Bloch Hamiltonian can be rewritten as
\begin{eqnarray}\label{eqn:SIH2}
{\cal H}(\bold k)&=&\biggr(t_1+2t_0\cos\frac{k_1+k_2}{2}\cos\frac{k_1-k_2}{2}\biggr)\sigma_x+2t_0\sin\frac{k_1+k_2}{2}\cos\frac{k_1-k_2}{2}\sigma_y\nonumber\\
&&+2t_2\sin\phi\biggr[2\cos\frac{k_1+k_2}{2}\sin\frac{k_1-k_2}{2}-\sin(k_1-k_2)\biggr]\sigma_z.
\end{eqnarray}
Similar as the case in the square lattice model, the above Hamiltonian is invariant under the inversion transformation defined by $P=\hat P\otimes\hat R_{\rm 2D}$, with the parity operator $\hat P=\sigma_x$. This parity operator is different from that in the square lattice system studied in the main text, since in the honeycomb lattice this operator acts on the sublattice space. Note that the Bloch momentum $\bold k=(k_1,k_2)$ (for honeycomb lattice the TR and spatial inversion invariant momenta are given by $(k_1,k_2)=(n_1\pi,n_2\pi)$, not $(k_xa,k_ya)=(n_1\pi,n_2\pi)$). Again, at the four symmetric points $\{\bold \Lambda^{(i)}\}=\{(\Lambda^{(i)}_1,\Lambda^{(i)}_2)\}=\{(0,0), (0,\pi), (\pi,0), (\pi,\pi)\}$ we have that $[\hat P,{\cal H}(\bold \Lambda_i)]=0$, and the Bloch states $|u_{\pm}(\bold \Lambda_i)\rangle$ are also eigenstates of the parity operator $\hat P$, with eigenvalues $\xi^{(\pm)}_i=+1$ or $-1$. Using the Eq.~(2) in the main text we have
\begin{eqnarray}\label{eqn:honeycomb1}
(-1)^{\nu}&=&\prod_{i=1}^4\biggr(t_1+2t_0\cos\frac{\Lambda^{(i)}_1+\Lambda^{(i)}_2}{2}\cos\frac{\Lambda^{(i)}_1-\Lambda^{(i)}_2}{2}\biggr)\nonumber\\
&=&\mbox{sgn}\bigr[(t_1+2t_0)(t_1-2t_0)\bigr].
\end{eqnarray}
From the above formula we can find that $\nu=1$ when $|t_1|<2t_0$ and $\nu=0$ otherwise. On the other hand, to ensure the Hamiltonian~\eqref{eqn:SIH2} to have a finite bulk gap, one requires that $t_1\neq0$. Therefore the topological phase can be obtained in the parameter regime $|t_1|<2t_0$ and $t_1\neq0$. Note that for $t_1=0$ the bulk gap closes at two different points $(k_1,k_2)=(0,\pi)$ and $(\pi,0)$. Thus the Chern number does not change when $t_1$ varies from $t_1\gtrsim0$ to $t_1\lesssim0$, and we finally conclude that
\begin{eqnarray}\label{eqn:honeycomb2}
C_1= \left\{ \begin{array}{ll}
         \mbox{sgn}(\sin\phi), \ \ \ \mbox{for} \ -2t_0<t_1<2t_0 \ \mbox{and} \ t_1\neq0,\\
        0, \ \ \ \ \ \ \ \ \ \ \ \ \ \ \mbox{otherwise}.\\
        \end{array} \right.
\end{eqnarray}
In the realistic lattice model $t_1$ is always positive (or in the same sign as $t_0$), and then the critical point in the experiment can be obtained by tuning it to be $t_1=2t_0$. It is interesting that using the invariant defined by Eq.~(2) in the main text we have predicted here a new topological phase transition upon varying the parameter $t_1$.

Now we discuss the measurement, and we consider the second scheme introduced in the main text. Note that the parity operator is the pseudospin $\sigma_x$. The parity eigenvalues are the pseudospin eigenstates, and to measure the parity eigenvalues one can measure the pseudospin polarization along the $x$ direction (see below). The pseudospin polarization is defined by $p_s=(n_+-n_-)/(n_++n_-)$, where $n_\pm$ represent the atom densities in the pseudospin-$\pm$ states corresponding to $\sigma_x$. It is straightforward to know that
\begin{eqnarray}\label{eqn:honeycomb3}
(-1)^{\nu}&=&\prod_{i=1}^4\mbox{sgn}\bigr[p_s(\bold\Lambda^{(i)})\bigr].
\end{eqnarray}
Similar as the case in the square lattice model, one always has $p_s(\bold\Lambda^{(2)})=p_s(\bold\Lambda^{(3)})$. We then only need to measure $\mbox{sgn}\bigr[p_s(\bold\Lambda^{(1)})\bigr]$ and $\mbox{sgn}\bigr[p_s(\bold\Lambda^{(4)})\bigr]$. Note to evolve the momentum along the direction from $\bold\Lambda^{(1)}=(0,0)$ to $\bold\Lambda^{(4)}=(\pi,\pi)$, the external force is applied in the $x$ direction (refer to Fig.~\ref{lattice}). The results are completely similar as those shown in Fig.~2 (a) and (b) in the main text. When $0<t_1<2t_0$, in a whole Bloch oscillation the sign of the pseudospin polarization $p_s(\tau)$ reverses from $\mbox{sgn}[p_s(0)]=-1$ to $\mbox{sgn}[p_s(\tau)]=+1$ at $\tau=T_B/2$ and changes back to $\mbox{sgn}[p_s(\tau)]=-1$ when one Bloch oscillation is finished at $\tau=T_B$, implying that the system is in the topological regime. On the other hand, when tuning to $t_1>2t_0$, the sign of the polarization keeps unchanged during the Bloch oscillation, and the phase is trivial.

Finally, to measure the pseudospin-polarization corresponding to $\sigma_x$, one can shine the atoms with a plane wave laser propagating along $x$ direction in Fig.~\ref{lattice}. We consider the following two cases. First, the wave length of the laser equals the lattice constant of the honeycomb lattice; Second, the wave length of the laser is twice of the honeycomb lattice constant. We set that the frequency of the shining laser so that the coupling is resonant between the atomic state trapped in the optical lattice and some excited level. For the former case, the laser couples to the atomic states of the A and B sites in a unit cell with the same phase (or $2\pi$ difference), therefore it can excite the eigenstate of $\sigma_x$ with eigenvalue $+1$, which is an even superposition of A and B sites in a unit cell, to the excited level resonantly, but cannot excite the eigenstate of $\sigma_x$ with eigenvalue $-1$ which is an odd superposition of A and B sites in a unit cell (one can verify that the transition matrix vanishes). Accordingly, in the latter case, the laser couples to such two atomic states in a unit cell with a $\pi$ phase difference, therefore it can excite resonantly the eigenstate of $\sigma_x$ with the eigenvalue $-1$, but cannot excite the eigenstate of $\sigma_x$ with the eigenvalue $+1$. With this method we can detect the pseudospin polarization, and therefore the parity eigenstates corresponding to $\sigma_x$.



\end{document}